%% file: main.tex
\documentclass[conference]{IEEEtran}
\IEEEoverridecommandlockouts
\usepackage{cite}
\usepackage{amsmath,amssymb,amsfonts}
\usepackage{amsthm}
\usepackage{algorithmic}
\usepackage{graphicx}
\usepackage{textcomp}
\usepackage{xcolor}
\usepackage{dsfont}
\usepackage[hidelinks]{hyperref}
\usepackage{quantikz}

\begin{document}

\title{Perspectives on Utilization of Measurements in Quantum Algorithms}

\input{authors}

\maketitle

\begin{abstract}
Measurement is a fundamental operation in quantum computing and has many important use cases in quantum algorithms. This article provides a comprehensive overview of the basic measurement operations in quantum computing and represents a selected set of their applications in quantum algorithms. Our goal is to provide one of the first algorithmic overviews of measurement processes in quantum computing. From the quantum information-theoretical perspective, measurements are either a method to access the result of a quantum computation or a technique to modify a quantum state. We also identify measurement-based methods to solve quantum computational challenges, such as error mitigation and circuit cutting. We discuss three main categories of measurements: performing measurements in static quantum circuits, modifying the quantum state in dynamic quantum circuits via measurements, and addressing challenges in quantum computing with measurements. Based on the reviewed topics, the measurement operations are frequently not at the center of the quantum algorithm design. However, the most novel and error-prone quantum algorithms will likely require sophisticated measurement schemes. Thus, the central message of this article is to broaden the view of measurement operations and highlight their importance at every level of quantum algorithm design.
\end{abstract}


\input{sections/introduction}
\input{sections/background}
\input{sections/static}

\input{sections/dynamic}
\input{sections/challenges}

\input{sections/conclusion}

\bibliographystyle{IEEEtran}
\bibliography{ref.bib}

\end{document}

%% file: authors.tex
\author{
\IEEEauthorblockN{Valter Uotila}\thanks{Authors from University of Helsinki can be contacted via \texttt{first.last@helsinki.fi}}
\IEEEauthorblockA{\textit{University of Helsinki}\\
\textit{Aalto University}\\
}
\and
\IEEEauthorblockN{Ilmo Salmenperä}
\IEEEauthorblockA{\textit{University of Helsinki}\\
}
\and
\IEEEauthorblockN{Leo Becker}
\IEEEauthorblockA{\textit{University of Helsinki}\\
}
\and
\IEEEauthorblockN{Arianne Meijer - van de Griend}
\IEEEauthorblockA{\textit{University of Helsinki}\\
}
\and
\IEEEauthorblockN{Aakash Ravindra Shinde}
\IEEEauthorblockA{\textit{University of Helsinki}\\
}
\and
\IEEEauthorblockN{Jukka K. Nurminen}
\IEEEauthorblockA{\textit{University of Helsinki}\\
}
}

%% file: sections/introduction.tex
\section{Introduction}

Quantum computing is characterized by its fundamentally distinct approach to information processing. Its computational features arise from quantum mechanical phenomena such as superposition and entanglement, which can be used to solve particular problems exponentially or polynomially faster \cite{365700, Grover_1996}. The main reason for these theoretical quantum speedups is the clever utilization of superposition, entanglement, interference, and state preparation. One of the most crucial goals in the field is to develop quantum hardware and algorithms that would demonstrate quantum advantage in a practically relevant problem with broad consensus \cite{Schuld_Killoran_2022}. 

In addition to the previous features, quantum computing is also characterized by quantum measurements, which are performed at the end of every quantum circuit. We have identified that measurements often do not receive the attention they deserve in the design of quantum algorithms. Thus, we have collected crucial measurement-related applications and categorized them into three categories: static, dynamic, and challenges.

The quantum algorithm-level topics in this review demonstrate that quantum measurement is not just a final, passive readout method, but an active part of quantum computational pipelines. We highlight the algorithmic perspective, as we are unaware of other review studies that concentrate on measurement operations in the context of quantum algorithms. Some articles and books focus on quantum measurement theory \cite{Busch_Lahti_1996, Jacobs_2014}. Still, they leave out current quantum computing use cases and examples of how measurements are applied in practical algorithms, such as quantum machine learning. This article addresses several fundamental measurement-related questions: 
\begin{itemize}
    \item What is the current landscape of the types of measurements used in quantum algorithms?
    \item How do measurements affect quantum computation?
    \item What are the key applications where measurements are used as a computational tool?
\end{itemize}

To comprehensively understand measurement operations, we identify three higher-level categories: measurements in static quantum circuits, measurements in dynamic quantum circuits, and measurements designed to tackle specifically quantum computational problems, such as error mitigation. By measurements in static circuits, we mean those operations that collapse the complete state and extract classical information without continuing quantum computation. Most current quantum algorithms employ these types of measurements. Measurements in dynamic quantum circuits refer to those operations that affect only a part of the measured state and perform a post-measurement state transformation. After the measurement, quantum computation is continued, and the system's state is modified. The evolution of the post-measurement state often depends on the outcomes of these mid-circuit measurements. 

In many quantum algorithms, such as the Quantum Approximate Optimization Algorithm (QAOA), we evaluate the expectation value of a cost Hamiltonian that encodes the structure of the optimization problem. In these cases, the measurement operator directly reflects the structure of the real-world problem being solved. The first two categories of measurement applications share this feature; their measurement schemes are part of algorithms or applications. In contrast, the third category comprises quantum computational problems and measurement applications that are designed to address challenges within quantum computing itself.

Advanced measurement operations have been developed and utilized in quantum hardware \cite{de_Leon_Itoh_Kim_Mehta_Northup_Paik_Palmer_Samarth_Sangtawesin_Steuerman_2021} and in fault-tolerant quantum computing \cite{Shor_1997, Preskill_1997, Gottesman_1998}. Based on our review, the utilization of advanced measurement operations has not made a significant breakthrough in the algorithms closer to the end-user applications, such as quantum machine learning and optimization. Measurements in quantum computing are often treated as final readout steps. However, measurement operations become increasingly critical when quantum computing algorithms become more sophisticated. More precisely, we consider the following topics:
\begin{itemize}
    \item Basics of quantum measurement theory: positive operator-valued measure (POVM) measurements, projective (PVM) measurements, and mid-circuit measurements.
    \item Measurements in static circuits: observables encode optimization problems, measurement-related trainability issues, optimizing measurement operations for specific tasks, and how measurements in variational quantum algorithms differ from measurements in foundational quantum algorithms.
    \item Measurements in dynamic circuits: state preparation and quantum teleportation.
    \item Measurements for quantum computational challenges: Circuit cutting and error mitigation.
\end{itemize}




%% file: sections/background.tex
\section{Background on quantum measurement theory}

This section introduces the background of quantum measurement theory. Considering the overview of the theoretical landscape, we connect the two fundamental quantum state definitions, pure and mixed states, and measurements to give a new perspective on quantum measurement theory. The correspondence is presented in \autoref{table:conceptual_correspondence_between_objects_measurements}, but it should not be considered to show mathematical equivalence.

\input{tables/QC_vs_measurements}

Quantum computing evolves as a closed quantum system following the unitary evolution \cite{Nielsen_Chuang_2010}. The system must be measured to obtain an outcome from a quantum computational process. Measurement breaks the closed unitary evolution, but it is the only way to access information from the system.

\input{sections/background/PVM}
\input{sections/background/POVM}
\input{sections/background/mid-circuit}

%% file: tables/QC_vs_measurements.tex
\begin{table*}[t]
\centering
\begin{tabular}{|c|c|c|c|}
\hline
\begin{tabular}{c}
\textbf{Measurement} \\
\textbf{theory}
\end{tabular} &
\textbf{Properties} &
\begin{tabular}{c}
\textbf{Quantum} \\
\textbf{states}
\end{tabular} &
\textbf{Properties} \\
\hline
\begin{tabular}{c}
\textbf{PVM} \\
\textbf{measurements}
\end{tabular} &
\begin{tabular}{c}
Idempotent: $P_i^2 = P_i$ \\
Orthogonality: $P_i P_j = \delta_{ij}P_i$ \\
Completeness: $\sum_i P_i = I$
\end{tabular} &
\begin{tabular}{c}
\textbf{Pure} \\
\textbf{states}
\end{tabular} &
\begin{tabular}{c}
Idempotent: $\rho^2 = \rho$ \\
Projection onto 1D subspace \\
Normalization: $\text{Tr}(\rho) = 1$
\end{tabular} \\
\hline
\begin{tabular}{c}
\textbf{POVM} \\
\textbf{measurements}
\end{tabular} &
\begin{tabular}{c}
Positive operators $E_i \geq 0$ \\
Completeness: $\sum_i E_i = I$ \\
Not necessarily orthogonal \\
Generalization of PVMs
\end{tabular} &
\begin{tabular}{c}
\textbf{Mixed} \\
\textbf{states}
\end{tabular} &
\begin{tabular}{c}
Positive operators $\rho \geq 0$ \\
Normalization: $\text{Tr}(\rho) = 1$ \\
Not necessarily pure \\
Generalization of pure states
\end{tabular} \\
\hline
\end{tabular}
\vspace{1em}
\caption{Conceptual similarity between measurement operations and quantum states}
\label{table:conceptual_correspondence_between_objects_measurements}
\end{table*}

%% file: sections/background/PVM.tex
\subsection{Projective (PVM) measurements}

Projection-Valued Measures (PVMs) are the most practical category of measurements \cite{Nielsen_Chuang_2010}. PVMs are usually called projective measurements. Projective measurements are defined as observables, which are Hermitian (self-adjoint) operators acting on the observed state space. If $O$ is an observable, the spectral decomposition for this operator gives us $O = \sum_{\lambda}\lambda P_{\lambda}$ in terms of eigenvalues $\lambda$ and eigenvectors $P_{\lambda}$ of $O$. We say that $P_{\lambda}$ is the projector onto the eigenspace of $O$ with eigenvalue $\lambda$. The possible results of this projective measurement are the eigenvalues $\lambda$ of the observable $O$. For a given state $|\varphi \rangle$, the probability of measuring an eigenvalue $\lambda$ is defined by $p(\lambda) = \langle \varphi | P_{\lambda} | \varphi \rangle$.

The projectors $P_{\lambda}$ of $O$ satisfy the completeness relation
\begin{equation}\label{eq:completeness_pvm}
    \sum_{\lambda}P_{\lambda} = I,
\end{equation}
they are idempotent
\begin{equation}\label{eq:idempotence_pvm}
   P_{\lambda}^2 := P_{\lambda}P_{\lambda} = P_{\lambda},
\end{equation}
and they are also orthogonal in the sense that 
\begin{equation}\label{eq:orthogonality_pvm}
    P_{\lambda}P_{\lambda'} = 0, \text{ when } \lambda \neq \lambda'.
\end{equation}

We do not have to measure the entire space at once, which is a key property that enables mid-circuit measurements. When measuring only a subsystem, the post-measurement state update for a projective measurement is given by
\begin{equation}\label{eq:post_measurement_state_transformation}
\frac{P_{\lambda}|\varphi\rangle}{\sqrt{\langle\varphi|P_{\lambda}^{\dagger}P_{\lambda}|\varphi\rangle}} = \frac{P_{\lambda} |\varphi\rangle}{\sqrt{p(\lambda)}}.
\end{equation}

As pointed out in \cite{Nielsen_Chuang_2010}, one practical reason projective measurements are useful is that computing their expectation values is straightforward. Measuring the observable $O$ automatically produces the expectation value as
\begin{align}\label{eq:expectation_value}
    \mathds{E}[O] = \sum_{\lambda}\lambda p(\lambda) = \sum_{\lambda} \lambda \langle \varphi | P_{\lambda} | \varphi \rangle &= \langle \varphi | \sum_{\lambda}\lambda P_{\lambda} | \varphi \rangle \\ 
    &= \langle \varphi | O | \varphi \rangle.
\end{align}

An important set of observables can be constructed from Pauli matrices:
\begin{displaymath}
    \sigma_x = \begin{bmatrix} 0 & 1 \\ 1 & 0 \end{bmatrix}, \quad
    \sigma_y = \begin{bmatrix} 0 & -i \\ i & 0 \end{bmatrix}, \quad
    \sigma_z = \begin{bmatrix} 1 & 0 \\ 0 & -1 \end{bmatrix}.
\end{displaymath}
Each Pauli matrix is Hermitian and has eigenvalues $\pm 1$ with orthonormal eigenvectors. For example, the eigenvectors of $\sigma_z$ are $|0\rangle$ and $| 1\rangle$. The corresponding projectors for $\sigma_z$ are:
\begin{equation}\label{eq:computational_basis}    
P_{+z} = |0 \rangle \langle 0| = \frac{I + \sigma_z}{2}, \quad
P_{-z} = |1 \rangle \langle 1| = \frac{I - \sigma_z}{2}.
\end{equation}
The direct computation shows that $P_{\pm z}^2 = P_{\pm z}$ and $P_{+z}P_{-z} = P_{-z}P_{+z} = 0$, so $P_{\pm z}$ define a projective measurement by the definition.

Given a single qubit state $| \varphi \rangle = \alpha |0\rangle + \beta |1\rangle$ for $\alpha, \beta \in \mathds{C}$ and the previous projectors $P_{\pm z}$, we obtain 
\begin{align}\label{eq:measurement_in_computational_basis}
    \langle \varphi | 0 \rangle \langle 0 | \varphi \rangle = \overline{\alpha}\alpha = |\alpha|^2 \text{ and } 
    \langle \varphi | 1 \rangle \langle 1 | \varphi \rangle = \overline{\beta}\beta = |\beta|^2.
\end{align}
This is a measurement in the computational basis. Consequently, the probability of measuring $0$ is $|\alpha|^2$, and the probability of measuring $1$ is $|\beta|^2$. While the requirement that $|\alpha|^2 + |\beta|^2 = 1$ is often stated as part of the definition of a qubit, it necessarily follows from the properties of this measurement. This is also often the first measurement operation learned when studying quantum computing.

The previous reasoning naturally extends to multi-qubit systems with respect to their computational bases. The earlier approach also extends to measure in any orthonormal basis \cite{Wilde_2016} in the following sense. Let $\left\{ |\varphi_{i}\rangle \right\}_{i \in I}$ be an orthonormal basis set. Then the projectors are defined by $|\varphi_{i} \rangle \langle \varphi_{i}|$ for each $i \in I$. The probability measuring the state $|\varphi_i\rangle$ is given by $\langle \phi |\varphi_i\rangle \langle \varphi_i | \phi \rangle = |\langle \varphi_i | \phi \rangle|^2$ and by \autoref{eq:post_measurement_state_transformation} the post-measured state is
\begin{displaymath}
    \frac{|\varphi_i\rangle \langle \varphi_i | \phi \rangle}{|\langle \varphi_i | \phi \rangle|}.
\end{displaymath}

The last row in \autoref{table:conceptual_correspondence_between_objects_measurements} claims that projective measurements are conceptually linked to pure states. Recall that a pure state is described by a density matrix $\rho = |\varphi\rangle\langle\varphi|$. This density operator has the following properties. It is idempotent:
\begin{equation*}
    \rho^2 = |\varphi\rangle\langle\varphi|\varphi\rangle\langle\varphi| = \rho,
\end{equation*}
which corresponds to the idempotence of projective measurements in \autoref{eq:idempotence_pvm}. The pure state is a projection onto a one-dimensional subspace corresponding to orthogonality in \autoref{eq:orthogonality_pvm}. Also, $\mathrm{Tr}(\rho) = 1$, which can be viewed as a correspondence to the completeness \autoref{eq:completeness_pvm}.

%% file: sections/background/POVM.tex
\subsection{POVM measurements}\label{subsec:povm}

A Positive Operator-Valued Measure (POVM) measurement is a general formalism describing quantum measurements \cite{Nielsen_Chuang_2010, Wilde_2016}. Let $M$ be a set of measurement outcomes that can happen in the experiment. Let $\left\{E_m\right\}_{m \in M}$ be a set of measurement operators. If the quantum system is in the state $| \varphi \rangle$ before the measurement, then the probability that the measurement outcome $m$ occurs is
\begin{displaymath}
p(m) = \langle \varphi | E_{m}| \varphi \rangle.
\end{displaymath}

The measurement operators $\left\{E_m\right\}_{m \in M}$ satisfy the completeness equation
\begin{equation}\label{eq:completeness_povm}
    \sum_{m \in M} E_{m} = I.
\end{equation}
The completeness ensures that the measurement outcomes form a well-defined probability distribution
\begin{equation*}
    1 = \sum_{m \in M}p(m) = \sum_{m \in M}\langle \varphi | E_{m}| \varphi \rangle.
\end{equation*}
The operators $\left\{E_m\right\}_{m \in M}$ are also positive \cite{Nielsen_Chuang_2010}
\begin{equation}\label{eq:positiveness_povm}
    E_m \geq 0 \text{ for each } m \in M.
\end{equation}

POVM measurements are instrumental when the only interest is measurement statistics, not post-measurement states, which were defined for the projective measurements in \autoref{eq:post_measurement_state_transformation}. Nevertheless, we can compute the post-measurement states with POVMs in the following sense. It can be proved that given any set of POVM operators $\left\{E_m\right\}_{m \in M}$, there exists a set of measurement operators $\left\{ K_m \right\}_{m \in M}$ so that the measurements $\left\{ K_m \right\}_{m \in M}$ define the POVM measurements: $E_m = K^{\dag}_m K_m$. The decomposition $E_m = K^{\dag}_m K_m$ of POVMs is not necessarily unique. Different choices of $\left\{ K_m \right\}_{m \in M}$ that give the same POVM $\left\{E_m\right\}_{m \in M}$ will produce identical measurement statistics but potentially different post-measurement states.

Projective measurements are also POVM measurements because $P_{\lambda}^{\dag}P_{\lambda} = P_{\lambda}$ since the projections are Hermitian and orthogonal. As proved in \cite{Nielsen_Chuang_2010}, the definition of projective measurements is equivalent to the definition of POVM measurements when the computational model is equipped with the possibility of performing unitary transformations. Hence, in practice, projective measurements are enough. However, as our examples in this paper will demonstrate, it is often useful to have varying theoretical measurement tools to describe measurement processes.

Recall that density matrices are a formalism to describe mixed states. If $p_i$ is the probability of measuring the state $|\varphi_i\rangle$, then the density matrix is defined by
\begin{equation*}
    \rho := \sum_{i}p_i |\varphi_i\rangle\langle\varphi_i|.
\end{equation*}
If $\rho$ is a density matrix, then it is a positive semi-definite
\begin{equation}\label{eq:positiveness_density}
    \rho \geq 0,
\end{equation}
and it has trace one
\begin{equation}\label{eq:trace_one_density}
    \mathrm{Tr}(\rho) = 1.
\end{equation}

Considering \autoref{table:conceptual_correspondence_between_objects_measurements} and the claim that POVM measurements conceptually correspond to mixed states, we note that the positiveness criterion in \autoref{eq:positiveness_povm} corresponds to positiveness of density operators stated in \autoref{eq:positiveness_density}. Similarly, the completeness \autoref{eq:completeness_povm} corresponds to the fact that density operators have trace one, stated in \autoref{eq:trace_one_density}. As POVMs generalize PVMS, mixed states generalize pure states. This hierarchical similarity between the measurements and states might help comprehend the larger theoretical picture of quantum computing. Next, we introduce specific POVM measurements.

\subsubsection{SIC-POVMs}\label{subsec:sic-povms} To refine the POVM measurements, we define Symmetric Informationally Complete Positive Operator-Valued Measure (SIC-POVM) measurements. We first discuss the ''informationally complete'' (IC) part and then include the ''symmetry'' (S) part. Thus, we define two sets of measurements: IC-POVMs and SIC-POVMs. Both of these measurement schemes can be used to recover the complete state.

Informational completeness refers to the property of POVM measurements, which allows us to reconstruct states from measurement results completely. More formally, a POVM $\left\{E_m\right\}_{m \in M}$ is considered to be informationally complete if it spans the space of Hermitian operators \cite{DAriano_Perinotti_Sacchi_2004, Fischer_Dao_Tavernelli_Tacchino_2024}. This means that for every Hermitian operator $O$, there exists $w_m \in \mathds{R}$ for $m \in M$ such that
\begin{equation*}
    O = \sum_{m \in M}w_m E_m.
\end{equation*}
Then the expectation value of the operator $O$ becomes
\begin{equation*}
    \mathds{E}[O] = \sum_{m \in M}w_m \langle \varphi | E_m | \varphi \rangle = \sum_{m \in M}w_m p(m).
\end{equation*}
In other words, if $\left\{E_m\right\}_{m \in M}$ is an IC-POVM, then
\begin{equation*}
    \langle \varphi_0 |E_m|\varphi_0 \rangle = \langle \varphi_1 |E_m|\varphi_1 \rangle
\end{equation*}
for all $m \in M$ holds if and only if $|\varphi_0 \rangle = | \varphi_1 \rangle$. This means that the measurement probabilities uniquely define the states \cite{Filippov_Leahy_Rossi_Garcia_Perez_2023}. This is the key reason why IC-POVMs are useful: the measurement statistic can recover the precise state before the measurement.

Next, we briefly describe how to reconstruct states using IC-POVMs in a simple one-qubit case \cite{Filippov_Leahy_Rossi_Garcia_Perez_2023}. This will be later used in the tensor-network error mitigation algorithm in practice. Consider the following set of POVMs
\begin{align}\label{eq:common_single_qubit_povms}
    E_{+z} &= p_z |0\rangle\langle0|, &\quad E_{-z} &= p_z |1\rangle\langle1|, \\
    E_{\pm x} &= p_x | \pm \rangle \langle \pm |, &\quad E_{\pm y} &= p_y | \pm i \rangle \langle \pm i |,
\end{align}
where $p_{z}, p_{x}, p_{y} \in [0, 1]$ so that $p_{z} + p_{x} + p_{y} = 1$, i.e., the values form a probability distribution. Let $I = \left\{ \pm z, \pm x, \pm y\right\}$ be the indexing set. The previous set of POVMs is informationally complete because the dimension of the space $\mathrm{span}(\left\{ E_{k} \right\}_{k \in I})$ is $4$. 

The reconstruction of the state is done with so-called dual operators, which are given by
\begin{equation}\label{eq:dual_operator}
    D_{\pm \alpha} = \frac{I \pm \sigma_{\alpha}/p_{\alpha}}{2},
\end{equation}
where $\sigma_{\alpha}$ is the corresponding Pauli matrix, $p_{\alpha}$ is the probability and $\alpha \in \left\{x, y, z \right\}$.  With the dual operators, the reconstructed state $\rho$ has the following form
\begin{align}\label{eq:IC_POVM_reconstructed_state}
    \rho = \sum_{k \in I} \mathrm{Tr}(\rho E_k) D_k.
\end{align}
The construction also generalizes into multi-qubit systems \cite{Filippov_Leahy_Rossi_Garcia_Perez_2023}. Still, the IC-POVM measurements are performed locally at each qubit, which does not require finding the IC-POVM for the whole system. For instance, Qiskit POVM-Toolbox \cite{Fischer_Dao_Tavernelli_Tacchino_2024} implements the state reconstruction using the previous IC-POVM in \autoref{eq:common_single_qubit_povms}.

For a POVM to span the entire Hilbert space $\mathcal{H}$ of dimension $d = \dim(\mathcal{H})$, it must contain at least $d^2$ elements. If it contains exactly $d^2$ elements, the IC-POVM is called minimal. Moreover, if the elements of a minimal IC-POVM satisfy the condition
\begin{displaymath}
    \mathrm{Tr}(E_iE_j) = \frac{1}{d + 1}, \text{ when } i \neq j,
\end{displaymath}
then the POVM is called a symmetric informationally complete POVM (SIC-POVM) \cite{Renes_Blume_Kohout_Scott_Caves_2004}. It remains an open question whether SIC-POVM always exists.

For instance, the previous example of developing IC-POVM for a single qubit was not minimal because it contained six elements, while four elements are enough. A SIC-POVM for a single qubit can be defined with the following four operators \cite{Renes_Blume_Kohout_Scott_Caves_2004}
\begin{align*}
    E_1 &= |0\rangle, \quad E_2 = \frac{1}{\sqrt{3}}|0\rangle + \sqrt{\frac{2}{3}}|1\rangle, \\
    E_2 &= \frac{1}{\sqrt{3}}|0\rangle + \sqrt{\frac{2}{3}}e^{i \frac{2\pi}{3}}|1\rangle \\
    E_3 &= \frac{1}{\sqrt{3}}|0\rangle + \sqrt{\frac{2}{3}}e^{i \frac{4\pi}{3}}|1\rangle.
\end{align*}
In theory, SIC-POVMs offer more efficient state reconstruction due to their minimality compared to IC-POVMs.

%% file: sections/background/mid-circuit.tex
\subsection{Mid-circuit measurements}\label{subsec:mid_circuit_measurements}

Mid-circuit measurements are measurement operations that are performed at intermediate stages within a quantum circuit \cite{Rudinger_Ribeill_Govia_Ware_Nielsen_Young_Ohki_Blume_Kohout_Proctor_2021} instead of at the end of the circuit. Mid-circuit measurements produce classical information that can be used to control quantum operations in the circuit. For example, the measurement outcome can be used to control if a specific gate is applied to a qubit that has not yet been measured.

Theoretically, mid-circuit measurements are modeled using so-called quantum instruments, which are special quantum channels \cite{Davies_Lewis_1970, Wilde_2016}. Quantum channel formalism is relatively technical and requires additional background, so we instead present the definition of mid-circuit measurements in terms of the quantum measurement theory presented in this section.

Let us assume that an observer measures with a fixed outcome $m$ using a projective measurement operator $P_m$. Based on \autoref{eq:post_measurement_state_transformation}, the post-measurement state transformation is defined
\begin{equation*}
    |\varphi\rangle\mapsto \frac{P_m|\varphi\rangle}{\sqrt{p(m)}}.
\end{equation*}
The system can be designed to store the measurement outcome in a classical register. Then the system becomes
\begin{equation*}
    \frac{P_m|\varphi\rangle}{\sqrt{p(m)}} \otimes |m\rangle.
\end{equation*}
We can furthermore perform conditional operations which depend on the classical value $m$. We will later introduce applications for mid-circuit measurements.


%% file: sections/static.tex
\section{Measurements in static quantum circuits}

Static quantum circuits form the standard approach for how measurements are most often used in quantum computing~\cite{fang2023dynamicquantumcircuitcompilation}. In this class of circuits, the algorithms are composed of sequences of quantum gates and measurements at the end of the circuits. This section will outline how measurements are used in these quantum algorithms and examine the quantum computational consequences they produce.

\subsection{Measurement in fundamental quantum algorithms}
Measurement operations are one of the key features in the fundamental quantum algorithms \cite{jordan2025quantum}, among which Grover's and Shor's algorithms are the most famous \cite{Grover_1996, Shor_1997}. In Shor's algorithm, the Quantum Fourier Transform (QFT) is applied, and measurement collapses the state to a classical outcome used to infer periodicity. In Grover's algorithms, after multiple amplitude amplification steps, measuring the final state yields the marked item with high probability. In all these cases, measurement is used to extract classical results from the quantum state, and the algorithm's execution does not continue after measurement. Considering measurement as the final static readout method is the commonly held perspective on measurement operations.

The fundamental quantum algorithms often have a theoretical guarantee of the probability of measuring the correct answer. As many of these algorithms' results can be confirmed by testing whether the result is correct, like whether Shor's algorithm managed to factorize the number correctly, it will be easy to run the algorithm until a correct answer is found. The number of shots required for a correct answer will depend on the quality of the hardware in question.


\subsection{Variational quantum computing}
Measurement schemes in variational quantum algorithms have a different role compared to the many fundamental quantum algorithms \cite{Cerezo_Arrasmith_Babbush_Benjamin_Endo_Fujii_McClean_Mitarai_Yuan_Cincio_et_al_2021}. In variational algorithms, the circuit is iteratively executed with fixed parameters, and the parameters are classically tuned so that a certain objective or loss is minimized. Because the circuit is constantly measured during the hybrid quantum-classical optimization loop, the measurement is not the final result but rather a feedback of the current performance. Also, there is often no theoretical guarantee that we would measure the correct result compared to the fundamental quantum algorithms. Hence, the measurement operation has a conceptually different role in variational algorithms compared to the measurements in fundamental quantum algorithms.

The most common machine learning and optimization loss functions use measurements to evaluate observables. The quantum state is evolved using a parametrized circuit $U(\theta)$, where $\theta$ is a vector containing parameters that alter the behavior of the model. This leads to a definition for a general form of loss functions:
\begin{equation}\label{eq:loss_function}
\mathcal{L}_\theta(\rho, O) = \mathrm{Tr}\big(U(\theta) \, \rho \, U^\dagger(\theta) \, O\big),
\end{equation}
where $\rho$ is the state before the parametrized circuit is applied, and $O$ is a Hermitian observable. Specific loss functions can be constructed by varying $U(\theta)$, $\rho$, and $O$ in \autoref{eq:loss_function}. The observable $O$ is of special interest for this work, as it corresponds to how the quantum state is measured.

In some quantum machine learning models, there is freedom to modify the measurement operations \cite{Bowles_Ahmed_Schuld_2024, PhysRevA.111.042420}. For example, the position of the measured qubit is not necessarily fixed. There are relatively few studies on the effect of these types of measurements on quantum machine learning models and on understanding how they affect the performance, learnability, and quality of the results. Additionally, there is even more room for improvement when considering that we can execute multiple models in parallel and perform varying measurement schemes on otherwise similar circuits. For example, an example of this type of multi-circuit training is \cite{Lorenz_Pearson_Meichanetzidis_Kartsaklis_Coecke_2023}, but the authors do not consider modifying the measurements.

\subsubsection{Optimization problems and measurements}

Optimization is one of the key areas where quantum computing is believed to provide value in the future \cite{Abbas_Ambainis_Augustino_Bartschi_Buhrman_Coffrin_Cortiana_Dunjko_Egger_Elmegreen_et_al_2024}, and many quantum optimization methods are fundamentally variational algorithms. Some of the most common quantum algorithms to perform quantum optimization are QAOA \cite{Farhi_Goldstone_Gutmann_2014} and VQE \cite{Peruzzo_McClean_Shadbolt_Yung_Zhou_Love_Aspuru_Guzik_OBrien_2014}. Both algorithms utilize a problem-dependent measurement scheme, which is crucial to their success. In this section, we present a simple example illustrating how the measurement, specifically the Hamiltonian, encodes the problem's solution landscape and how it can be utilized to identify the correct solution using quantum-classical variational optimization strategies. The qubit requirements of quantum optimization algorithms can be reduced by considering measurements in multiple bases \cite{PhysRevResearch.4.033142, Sciorilli_Borges_Patti_Garcia_Martin_Camilo_Anandkumar_Aolita_2025}.

Let $z_1$ and $z_2$ be two spin variables, i.e., $z_1, z_2 \in \left\{ -1, 1 \right\}$. The optimization problem in this example is to minimize $z_1 + z_2$. The question is, how does this look in terms of measurements? Obviously, the solution is $z_1 = -1$ and $z_2 = -1$ so that the sum $z_1 + z_2 = -2$.

In the standard approach, we prepare a so-called cost Hamiltonian, i.e., a cost objective, which encodes the optimization problem. We aim to minimize the value of this objective function. Countless optimization problems can be encoded with Hamiltonians \cite{Lucas_2014}. This is usually done by utilizing the Pauli-$Z$ matrix $\sigma_z$. As pointed out earlier, the $\sigma_z$ matrix has eigenvalues $\pm 1$, which match with the spin variables. This motivates the idea that the variable $z_1$ is encoded with $\sigma_z^{1}:= \sigma_z \otimes I$ acting on the first qubit and $z_2$ is encoded with $\sigma_z^{2}:= I \otimes \sigma_z$ acting on the second qubit. Writing out the calculations, we can see the following
\begin{align*}
    z_1 + z_2 \mapsto \sigma_z^{1} + \sigma_z^{2} = \sigma_z \otimes I +  I \otimes \sigma_z 
    = \begin{bmatrix}
    \mathbf{2} & 0 & 0 & 0 \\
    0 & \mathbf{0} & 0 & 0 \\
    0 & 0 & \mathbf{0} & 0 \\
    0 & 0 & 0 & \mathbf{-2}
    \end{bmatrix}.
\end{align*}
The diagonal of the previous matrix encodes all the possible combinations that the objective function $z_1 + z_2$ can achieve with the spin variables. This is one of the reasons why it is convenient to use Hamiltonians to encode optimization problems.

Next, we prepare a state vector $|\varphi\rangle:= [\alpha_1, \alpha_2, \alpha_3, \alpha_4]^{\dag}$ and measure this state using the projective measurement based on the Hamiltonian:
\begin{align*}
& \langle \varphi |(\sigma_z^{1} + \sigma_z^{2})|\varphi\rangle \\
& = \begin{bmatrix}
\overline{\alpha_1} & \overline{\alpha_2} & \overline{\alpha_3} & \overline{\alpha_4}
\end{bmatrix}
\begin{bmatrix}
\mathbf{2} & 0 & 0 & 0 \\
0 & \mathbf{0} & 0 & 0 \\
0 & 0 & \mathbf{0} & 0 \\
0 & 0 & 0 & \mathbf{-2}
\end{bmatrix}
\begin{bmatrix}
\alpha_1 \\
\alpha_2 \\
\alpha_3 \\
\alpha_4
\end{bmatrix} \\ 
&= 2 \overline{\alpha_1}\alpha_1 + 0 \overline{\alpha_2}\alpha_2 + 0 \overline{\alpha_3}\alpha_3 - 2 \overline{\alpha_4}\alpha_4 \\ 
&= 2|\alpha_1|^2 - 2|\alpha_4|^2.
\end{align*}
As discussed in the projective measurements subsection, this is the expectation value of the observable $\sigma_z^{1} + \sigma_z^{2}$. Since we want to minimize the sum of the variables, $|\alpha_4|^2$ should be maximized while $|\alpha_1|^2$ should be minimized. In practice, this process is implemented by using parametrized quantum circuits that prepare states $|\varphi\rangle$. Then, there is a high probability of measuring the expectation value $-2$, which is also the optimal value for the objective function. Since the $\alpha_i$ values are amplitudes in a state, they have to satisfy $|\alpha_1|^2 + |\alpha_2|^2 + |\alpha_3|^2 + |\alpha_4|^2 = 1$, which leads to the fact that an optimal point solving our optimization problem is a state for which $|\alpha_4|^2 = 1$. Moreover, the state for which $|\alpha_4|^2 = 1$, is clearly $[0,0,0,1]^{\dag}$. The value $1$ is in the index $4$, whose binary representation is $11$ and whose corresponding spin-solutions are $z_1 = z_2 = -1$, which is the solution to the simple optimization problem.

\subsubsection{Non-linear activation functions via measurements}

For classical machine learning, non-linear activation functions are crucial for the models to learn, generalize, and be sufficiently expressive. With non-linear activation functions and in certain theoretical settings, classical machine learning models are able to approximate arbitrary functions \cite{Hornik_1991}, i.e., they are universal approximations. Universal approximation is also a property that is desired from the quantum machine learning models, and part of the expressiveness that enables this comes from the measurement operations. 

One of the first and simplest classical machine learning models is a one-layer perceptron. Next, we consider a one-layer perceptron with $n$ input neurons and a single output neuron. The input for the layer is a real-valued vector $x := (x_1, \ldots, x_n)$ whose inner product with the weight vector $w:=(w_1, \ldots, w_n)^{\top}$ is computed. Then, the constant bias term $b \in \mathds{R}$ is added to the inner product. Assuming that $f \colon \mathds{R} \to K$ is an activation function mapping to a compact set $K$, the output from the perceptron is $f(x \cdot w + b)$. 

In quantum machine learning, measurements have been used to implement non-linear activation functions in a one-layer perceptron \cite{Daskin_2018, Cao_Guerreschi_Aspuru_Guzik_2017, Tacchino_Macchiavello_Gerace_Bajoni_2019}. Since the quantum state turns into classical data after measurement, measuring between layers imposes a challenge in implementing multilayer perceptions with non-linear activation functions based on measurements. Thus, the non-linear activation function should be approximated during the computation using linear operations, and only in the end should the state be measured. To briefly review the method in \cite{Maronese_Destri_Prati_2022}, the technique is implemented so that the powers of $ x\cdot w + b$ are stored in the amplitudes of a quantum state. Then, the Taylor expansion of $f(x \cdot w + b)$ of degree $d$ is constructed for a selected activation function $f$ utilizing the previously computed powers. This expansion gives us the quantum approximation of the non-linear activation function.

\subsubsection{Global observables lead to barren plateaus}

Another interesting measurement-related effect in quantum machine learning and optimization is barren plateaus, i.e., vanishing gradients \cite{McClean_Boixo_Smelyanskiy_Babbush_Neven_2018}. They arise from the quantum mechanical nature of the models, meaning that only models whose gradients are estimated with measurements suffer from barren plateaus. In classical machine learning, transformers suffered from trainability problems, and deep transformer models cannot be trained without residual connections \cite{Dong_Cordonnier_Loukas_2021, He_Zhang_Ren_Sun_2016}. There are interesting and unexplored connections between the trainability of classical models and the trainability of quantum machine learning models.

Recent research \cite{Ragone_Bakalov_Sauvage_Kemper_Ortiz_Marrero_Larocca_Cerezo_2024, Cerezo_Larocca_Garcia_Martín_Diaz_Braccia_Fontana_Rudolph_Bermejo_Ijaz_Thanasilp_2024} has characterized barren plateau-related results for variational quantum algorithms and quantum machine learning. Informally, a barren plateau means that the optimization landscape becomes flat when the size of the quantum computational system grows. More formally, barren plateaus are the phenomenon of quantum machine learning models where the variance of the loss function's gradient decreases exponentially as a function of the qubits in the system. In practice, this means that we have to measure the circuit an exponentially increasing number of times before we have a sufficiently accurate estimation for the gradient. The exponential scaling of sampling makes it impractical for larger quantum machine learning models, diminishing the potential quantum speedups.

\begin{figure}[t]
    \centering
    \resizebox{.5\columnwidth}{!}{%
    \input{circuits/global_observable_example}
    }
    \caption{Quantum circuit for an $n$-qubit system with $RY(\theta_i)$ rotations followed by a global observable $Z^{\otimes n}$}
    \label{fig:global_observable_barren_plateau_example}
\end{figure}
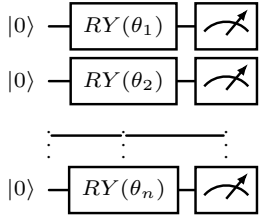

\begin{figure}[t]
    \centering
    \includegraphics[width=0.99\columnwidth]{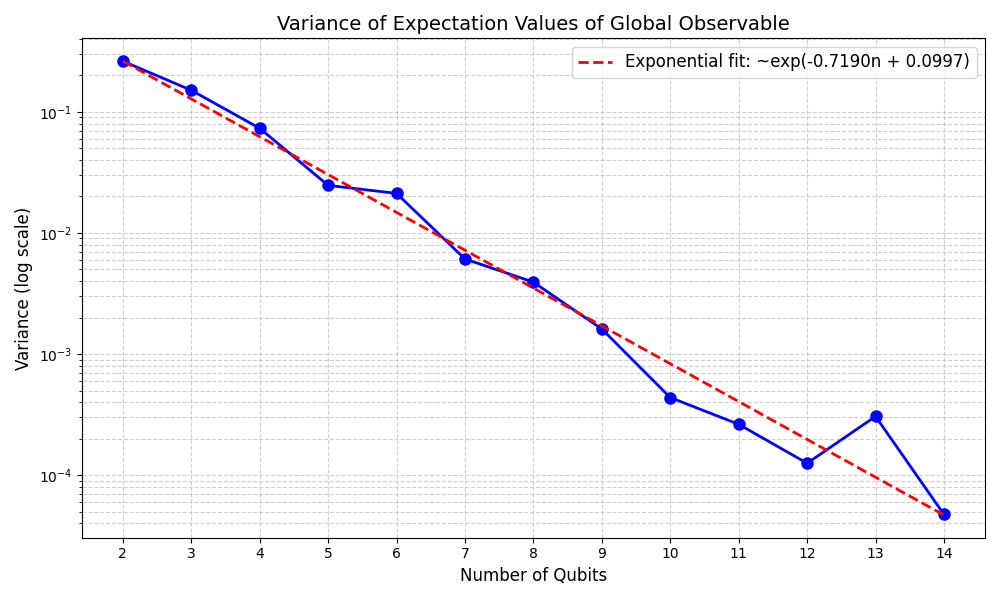}
    \caption{We computed the variances of expectation values from the circuit in \autoref{fig:global_observable_barren_plateau_example} for increasing numbers of qubits with randomly sampled parameters, demonstrating that the barren plateau arises from the use of the global observable $\bigotimes_{i = 1}^{n}\sigma_{z}^{i}$.}
    \label{fig:simple_circuit_barren_plateau}
\end{figure}

For certain types of quantum machine learning models, it was proved that the key sources for barren plateaus are expressive quantum circuits, entanglement, noise, and global observables \cite{Ragone_Bakalov_Sauvage_Kemper_Ortiz_Marrero_Larocca_Cerezo_2024}. For this article, the most relevant case is the global observables. \autoref{fig:global_observable_barren_plateau_example} demonstrates how global observables create barren plateaus even in straightforward setups. The studied model is based on the circuit in \autoref{fig:global_observable_barren_plateau_example}. It does not have entanglement, and it is not expressive since it has a single layer of $RY$ gates for any number of qubits. The measurement operation the circuit implements is the global observable of type $\bigotimes_{i = 1}^{n}\sigma_{z}^{i}$, where $n$ is the number of qubits in the model. We sample uniformly random variables for the parametrized $RY$ gates and execute the circuit on a noiseless simulator. This way, we can show that the discovered barren plateau phenomenon happens only because of the global observable. The results are presented in \autoref{fig:simple_circuit_barren_plateau}, which shows that the variance of the expectation values decreases exponentially as the size of the system increases.

While there is currently no fundamental solution for the barren plateau problem, the barren plateau-free models found to date are such that the measurable action of $U(\theta)$ can be characterized with only a polynomial cost for most parameter values $\theta$~\cite{Cerezo_Larocca_Garcia_Martín_Diaz_Braccia_Fontana_Rudolph_Bermejo_Ijaz_Thanasilp_2024}. This implies that barren plateau–free models exhibit a certain degree of classical simulatability. We want to emphasize that this does not apply in reverse, as shown in \autoref{fig:simple_circuit_barren_plateau}, where the model can clearly be simulated classically in polynomial time, but the model still suffers from the barren plateau problem. This leads us to realize that to perform meaningful measurements, we need to be deliberate in parameterizing quantum circuits. Some promising solutions for trainability issues in quantum machine learning have been developing cleverly structured circuits \cite{Kubler_Arrasmith_Cincio_Coles_2020, Larocca_Sauvage_Sbahi_Verdon_Coles_Cerezo_2022, Cerezo_Sone_Volkoff_Cincio_Coles_2021} or modifying the training pipeline \cite{Skolik_McClean_Mohseni_Van_Der_Smagt_Leib_2021}.

\subsubsection{Learned measurement operations}

While this section has mainly focused on variational algorithms and quantum machine learning, we also want to bring up the other direction: classical machine learning can improve the development of quantum computing algorithms in optimizing measurement operations.

We earlier described informationally complete POVM measurements in \autoref{subsec:povm}. Performing quantum measurements is always prone to errors and statistical inaccuracies, especially when larger systems are measured many times. In \cite{Garcia_Perez_Rossi_Sokolov_Tacchino_Barkoutsos_Mazzola_Tavernelli_Maniscalco_2021}, the authors propose a novel method to optimize IC-POVMs using classical machine learning so that certain statistical inaccuracies can be reduced. The method is also categorized as a variational algorithm, since it admits a variational subroutine.

The algorithm can be divided into three components. The first part is a hybrid quantum-classical sampling, which is applied to estimate the expectation values of observables. Let $O = \sum_{k} c_{k} \bigotimes_{j=1}^{N} \sigma^{(j)}_{k_j}$ be the decomposition of the observable $O$ in terms of Pauli matrices. Let $M$ be the set of measurement outcomes and $\left\{ E^{(i)}_m\right\}_{m \in M}$ be an IC-POVM, where $i$ denotes the qubit on which the operator acts. The method samples $S$ many probabilities $\left\{ p_{m_i} \right\}_{i = 1}^{S}$ where $m_i$ corresponds to the measurement outcomes from the POVM, and then computes the mean
\begin{equation}\label{eq:monte_carlo_method}
\bar{O} = \frac{1}{S} \sum_{s = 1}^{S}\omega_{m_{s}},
\end{equation}
where each term $\omega_{m_s} = \sum_{k}c_k\prod_{i}^{N}b^{(i)}_{k_im_i}$ \cite{Fischer_Dao_Tavernelli_Tacchino_2024}. The terms $c_k$ are obtained from the Pauli decomposition of $O$, and the terms $b^{(i)}_{k_im_i}$ are the coefficients when the Pauli operators are written in terms of the IC-POVM as
\begin{align*}
    \sigma^{(i)}_{k_i} = \sum_{m_i = 0}^{3}b_{k_im_i}^{(i)}E_{m_i}^{(i)}.
\end{align*}
By increasing the number of samples, one can estimate the expectation value $\mathds{E}[O] = \sum_{m}\omega_{m}p_{m}$ using \autoref{eq:monte_carlo_method} \cite{Garcia_Perez_Rossi_Sokolov_Tacchino_Barkoutsos_Mazzola_Tavernelli_Maniscalco_2021}.

In the second phase of the learned measurements algorithm, the method uses classical gradient descent to optimize the IC-POVM measurement operators $\left\{ E_m^{(i)}\right\}_{m \in M}$ to find an efficient set of measurements. The measurement operators are suitably parametrized, and the optimization target is to minimize the variance $\mathrm{Var}(\omega_m)$ for each measurement outcome $m$.

The final feature of the algorithm is that we do not need to wait until the end of the optimization process before we can start estimating the expectation values for $O$. Every step in the algorithm produces a valid set of IC-POVMs, which provide an expectation value estimate for $O$.

%% file: circuits/global_observable_example.tex
\resizebox{0.2\columnwidth}{!}{
\begin{quantikz}[row sep=0.1cm, column sep=0.2cm, font=\scriptsize]
\lstick{\ket{0}} & \gate{RY(\theta_1)} & \meter{} \\
\lstick{\ket{0}} & \gate{RY(\theta_2)} & \meter{} \\
\vdots & \vdots & \vdots \\
\lstick{\ket{0}} & \gate{RY(\theta_{n})} & \meter{}
\end{quantikz}
}

%% file: sections/dynamic.tex
\section{Measurements in dynamic quantum circuits}

This section outlines various use cases from the algorithmic point of view for dynamic quantum circuits and how they work. As shown in \autoref{subsec:mid_circuit_measurements}, dynamic quantum circuits modify the quantum state during the computation via mid-circuit measurements. Although initially introduced for error mitigation, this approach to quantum computing has since found valuable applications in quantum algorithm design.

An important feature of dynamic quantum circuits is that they can be created using static quantum circuits \cite{hong_22}. This process works by unconditionally applying the classically controlled operation and retaining only those outcomes where the measured qubit is found in the desired state at the end of the circuit (also called postselection). Although this static approach produces equivalent output probability distributions, mid-circuit measurements can sometimes be much more efficient than their static counterparts. For example, in \cite{baumer_24}, the authors showed that a quantum Fourier transform routine with measurements in the end can be optimized drastically using mid-circuit measurements from $\mathcal{O}(n^2)$ two-qubit gates to $\mathcal{O}(n)$ classically controlled one-qubit operations. In addition to containing fewer gates, this approach also does not require entangling operations, making it very simple to compile.

\subsection{State preparation using dynamic quantum circuits}

Dynamic quantum circuits can be used in an efficient state preparation. The motivation for the usage of mid-circuit measurements in these state preparation algorithms is to reduce circuit depth while accurately preparing the desired quantum state. A simple example is presented in \cite{baumer_24_2}, where the authors transform the GHZ-state preparation circuit from being linearly dependent in terms of qubits into a constant-depth circuit. A more complex example is \cite{fossfeig2023}, where the authors use dynamic quantum circuits to prepare long-range entangled topological states beyond the analytical upper limit of fidelity for static quantum circuits. 

While the most essential use case is the creation of more efficient state preparation circuits for algorithms, mid-circuit measurements can also be useful due to their ability to represent diverse methods of state construction. For instance, in \cite{wang2024}, the authors use a projective measurement in the middle of the circuit to implement an arbitrary phase-encoding unitary. In this method, the state preparation starts with the creation of a superposition over $k$ qubits with an ancilla qubit in the $\ket{+}$ state:
\begin{align*}
\ket{\psi} = \ket{+} \otimes \left( \frac{1}{\sqrt{K}} \sum_k  \ket{k}\right).
\end{align*}

Next, we apply multi-controlled $RZ$-gates to apply desired phases onto each state using a unitary:

\begin{align*}
U_k = e^{-ix_k Z_a \otimes \ket{k}\bra{k}}
\end{align*}
The $U_k$ can be used to modify the initial state to
\begin{align*}
\ket{\Psi} &= \prod_k U_k \ket{\psi} \\
&= \frac{1}{\sqrt{K}}\sum_k \left( \frac{e^{-ix_k}\ket{0} + e^{-ix_k}\ket{1}} {\sqrt{2}} \otimes \ket{k} \right).
\end{align*}
Projecting the ancilla qubit onto the state $\ket{-}$ via a mid-circuit measurement and filtering out incorrect outcomes allows us to prepare the desired output state:
\begin{align*}
\ket{\Psi_{\text{out}}} = \sum_k \sin x_k \ket{k} \approx \sum_k x_k \ket{k}
\end{align*}
for small values of $x_k$, since then $\sin x_k \approx x_k$. This shows how a mid-circuit measurement can construct an arbitrary phase-encoding unitary. This method allows us to conceptualize the algorithm more easily. 


\subsection{State and gate teleportation}

Quantum state and gate teleportation algorithms are foundational quantum computational techniques to transfer an exact state between two parties using an entangled pair and classical communication \cite{Gottesman_Chuang_1999, Nielsen_Chuang_2010}. Quantum teleportation is a quantum computational primitive that is applied in measurement-based quantum computing \cite{Raussendorf_Briegel_2001}, and one can implement fault-tolerant quantum computing based on the gate teleportation scheme \cite{Gottesman_Chuang_1999}. The quantum teleportation schemes utilize measurements without collapsing the whole computational state. These algorithms rely on mid-circuit measurements in distributed systems, although the teleportation literature does not necessarily refer to these as mid-circuit measurements.

\begin{figure}[t]
    \centering
    \resizebox{\columnwidth}{!}{%
    \input{circuits/single_gate_teleportation}
    }
    \caption{Circuit that implements gate teleportation through gate $U$ with the correction operation $C_{xy}$ where the correction operation $C_{xy}$ depends on the gate $U$ and $x$ and $y$ depend on the Bell measurement results from the two measured qubits.}
    \label{fig:single_gate_teleportation}
\end{figure}
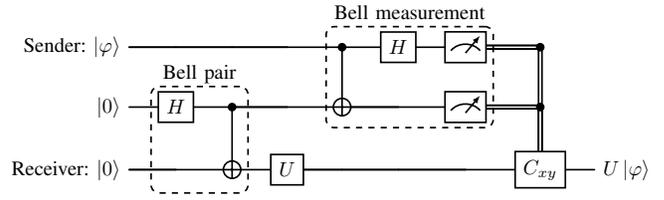

We briefly describe two main quantum teleportation algorithms: state and gate teleportation. In gate teleportation, we apply a unitary $U$ to the transferred state, whereas state teleportation transfers the state without modifying it. The circuit diagram in \autoref{fig:single_gate_teleportation} describes the gate teleportation scheme.

The state teleportation can be viewed as a special case of gate teleportation where we choose $U = I$ as the identity \cite{10821184}. In the case of state teleportation, the correction operation $C_{xy}$ becomes classically controlled $X$ and $Z$ gates acting on the receiver's qubit. The information about the required correction operations that the receiver should perform is transferred over a classical channel.

The teleportation schemes utilize the so-called Bell measurement, which is a measurement scheme where CNOT and Hadamard gates are applied in reverse order compared to the Bell state preparation. \autoref{fig:single_gate_teleportation} visualizes the Bell measurement. Then, the state is measured on a computational basis. Depending on the measured bits, the necessary corrections are applied. The exact form of correction operations depends on the gate $U$. For example, \cite{10821184} describes the special case of how to construct the correction operations for Toffoli-gate teleportation.

%% file: circuits/single_gate_teleportation.tex
\begin{quantikz}
  \lstick{Sender: $\ket{\varphi}$} & \qw & \qw & \qw & \ctrl{1} \gategroup[2,steps=3,style={dashed, rounded corners, inner ysep=-1pt, inner xsep=0pt}]{Bell measurement} & \gate{H} & \meter{} & \cwbend{2} \setwiretype{c} \\
  \lstick{$\ket{0}$} & \gate{H} \gategroup[2,steps=2,style={dashed, rounded corners, inner ysep=-1pt, inner xsep=0pt}]{Bell pair} & \ctrl{1} & \qw & \targ{}  & \qw & \meter{} & \cwbend{1} \setwiretype{c} \\
  \lstick{Receiver: $\ket{0}$} & \qw & \targ{} & \gate{U} & \qw & \qw & \qw & \gate{C_{xy}} & \rstick{$U\ket{\varphi}$}
\end{quantikz}

%% file: sections/challenges.tex
\section{Measurements as solutions to challenges in NISQ devices}

In this section, we will present two challenges that arise from quantum computing and address them with advanced measurement schemes. We identify this as a separate class from the previous static and dynamic measurement schemes because these problems are inherently quantum computational. The first problem is that quantum computing hardware does not scale to deep and wide circuits. The second problem is to address the error on the current hardware.

\subsection{Quantum hardware scalability problem}

Circuit cutting (or circuit knitting) is a distributed quantum computing method to execute a large circuit in multiple quantum computers with the cost of an exponential sampling overhead \cite{Bravyi_Smith_Smolin_2016, Eddins_Motta_Gujarati_Bravyi_Mezzacapo_Hadfield_Sheldon_2022, Brenner_Piveteau_Sutter_2023, Piveteau_Sutter_2024}. The sampling overhead is the number of shots to
achieve a fixed accuracy. Circuit cutting is beneficial when a large circuit does not fit into the available quantum computers, but it can be divided into smaller circuits that can be executed separately on the available devices. The results from these subexperiments are combined, approximating the original circuit's result.

Circuit cutting is divided into two methods: wire cutting and gate cutting. One of the first circuit cutting techniques was developed for wire cutting \cite{Peng_Harrow_Ozols_Wu_2020}, and Qiskit implements a toolbox to perform wire and gate cutting in practice \cite{qiskit-addon-cutting}. More advanced wire and gate-cutting methods can be realized by allowing classical communication between the parties. Depending on the form of classical communication, this creates a model with no classical communication, a model with one-way classical communication, or a model with two-way classical communication. Thus, these methods can be implemented and improved with mid-circuit measurements. Next, we focus on wire and gate cutting and how they utilize measurements to cut and distribute the circuits.

\subsubsection{Wire cutting}
\begin{figure}[t]
    \centering
     \resizebox{0.8\columnwidth}{!}{%
    \input{circuits/wire_cutting_example}
    }
    \caption{Conceptual idea behind wire cutting: the selected wire is cut, and the outcome is estimated on separate devices with suitable measurements and state preparations}
    \label{fig:wire_cutting_example}
\end{figure}
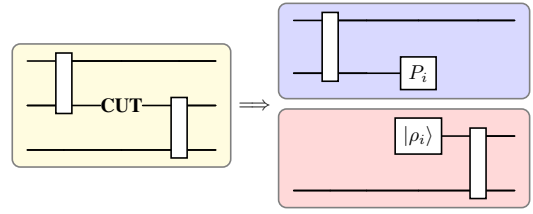

The idea of wire cutting is conceptually presented in \autoref{fig:wire_cutting_example}. The example of wire cutting we chose to present here highlights the role of measurements \cite{Peng_Harrow_Ozols_Wu_2020, Mitarai_Fujii_2021, Brenner_Piveteau_Sutter_2023} but does not utilize classical communication. In this example, wire cutting is built on projective measurements, states, and coefficients listed in \autoref{table:wire_cutting}.

\input{tables/wire_cutting_example}

The wire-cutting algorithm prepares a measurement scheme which is closely similar to the projective measurements already presented in \autoref{eq:common_single_qubit_povms}. This set has been extended by including the identity observables ($i = 1, 2$). Then, the reconstructed density operator is obtained with
\begin{equation}\label{eq:wire_cutting_decomposition}
\rho = \frac{1}{2} \sum_{i=1}^{8} c_i\mathrm{Tr}(P_i \rho)\rho_{i},
\end{equation}
where the values are listed in \autoref{table:wire_cutting}.

Now \autoref{eq:wire_cutting_decomposition} provides us a method to express the identity operator at the place where we want to perform the cut:
\begin{equation*}
    \input{circuits/wire_cutting_identity}
\end{equation*}
Every such cut produces eight circuit pairs, which can be executed separately. The separate measurement results are classically post-processed to finalize the complete circuit execution. Unfortunately, classical computational complexity grows exponentially with the number of performed cuts since we are necessarily required to simulate quantum computation at the positions where the cuts are performed.

Wire cutting methods with classical communication were further developed in \cite{Lowe_Medvidovic_Hayes_ORiordan_Bromley_Arrazola_Killoran_2023, Brenner_Piveteau_Sutter_2023}. Wire cutting can also be realized as a result of entanglement forging \cite{Eddins_Motta_Gujarati_Bravyi_Mezzacapo_Hadfield_Sheldon_2022} or quasiprobability distributions \cite{Brenner_Piveteau_Sutter_2023}. Next, we discuss the quasiprobability distributions-based method in the context of gate cutting.

\subsubsection{Gate cutting}

\begin{figure}[t]
    \centering
    \resizebox{0.8\columnwidth}{!}{%
    \input{circuits/gate_cutting_example}
     }
    \caption{Conceptual idea behind gate cutting: the green gates are cut, and the parts are executed on separate devices}
    \label{fig:gate_cutting_example}
\end{figure}
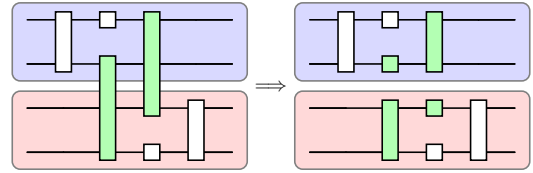

The conceptual idea behind gate cutting is visualized in \autoref{fig:gate_cutting_example}. The gate cutting was proposed in \cite{Mitarai_Fujii_2021, Mitarai_Fujii_2021_2} as a continuation of the wire cutting. The theory was further developed in \cite{Piveteau_Sutter_2024} to include classical communication to decrease the sampling overhead.

The key idea is to utilize quasiprobability distributions. If the gate $U$ is spanned across multiple quantum devices, it is expressed using a quasiprobability decomposition as
\begin{equation*}
    U = \sum_{i}a_i \mathcal{F}_i,
\end{equation*}
where $\mathcal{F}_i$ are operations supported by the devices and $a_i \in \mathds{R}$ are real numbers which might be negative. The operations $\mathcal{F}_i$ are not necessarily unitary, and they can also include measurements, as the following example demonstrates \cite{qiskit-addon-cutting, Mitarai_Fujii_2021}.

We perform gate cutting for a parametrized two-qubit gate $RZZ(\theta)$. The quasiprobability decomposition becomes
\begin{align*}
    RZZ(\theta) = \sum_{i = 1}^{6} a_i \mathcal{F}_i,
\end{align*}
where the elements of this sum are listed in \autoref{table:gate_cutting_example}. The algorithm produces six circuits for each $i = 1, \ldots, 6$. When the non-local gate is cut, the algorithm performs the first operation in the tensor product in $\mathcal{F}_i$ on the first qubit and the second operation on the second qubit.

\input{tables/gate_cutting_example}

As the example demonstrates, the projective measurements $P_z$ are crucial in this quasiprobability decomposition example. We are not decomposing in terms of unitaries but in terms of quantum channels. Quasiprobability distributions and decompositions have also been used in error mitigation \cite{Temme_Bravyi_Gambetta_2017, Endo_Benjamin_Li_2018, Kandala_Temme_Corcoles_Mezzacapo_Chow_Gambetta_2019, Piveteau_Sutter_Bravyi_Gambetta_Temme_2021, Piveteau_Sutter_Woerner_2022}, estimating outcome probabilities \cite{Pashayan_Wallman_Bartlett_2015}, and to prepare magic states \cite{Howard_Campbell_2017, Seddon_Campbell_2019, Heinrich_Gross_2019, Seddon_Regula_Pashayan_Ouyang_Campbell_2021}. 

In the gate cutting models, the effect of the non-local gate is simulated on a classical computer. The gate cutting can also be performed using only quantum computing resources to execute the non-local gate locally on an additional quantum computer using quantum and classical communication and the gate teleportation scheme \cite{10821184}. This creates no classical overhead since no classical resources are used.

Both gate and wire cutting can be more efficient with classical communication, i.e., if we utilize mid-circuit measurements and shared entanglement \cite{Brenner_Piveteau_Sutter_2023, Piveteau_Sutter_2024}. For example, without classical communication, the optimal wire cutting has complexity $4^{2n}$, where $n$ is the number of cuts \cite{Brenner_Piveteau_Sutter_2023}. Classical communication reduces the overhead to $(2^{n + 1} - 1)^2$. Surprisingly, the position of the cuts does not affect complexity, and these complexities were proved to be optimal for wire cuts \cite{Brenner_Piveteau_Sutter_2023}. For gate cutting, the classical overhead can be reduced from $9^{n}$ to $4^{n}$ if $n$ CNOT-gates are cut.

\subsection{Quantum hardware error problem}

Various measurement schemes can be employed in novel ways in quantum error mitigation. For example, mid-circuit measurements were applied for error mitigation in \cite{Botelho_Glos_Kundu_Miszczak_Salehi_Zimboras_2022} and quasi-probability distributions in \cite{Temme_Bravyi_Gambetta_2017}. In this subsection, we briefly focus on the tensor network error mitigation algorithm \cite{Filippov_Leahy_Rossi_Garcia_Perez_2023}, which utilizes IC-POVMs and cancels noise by applying noise-suppressing quantum channels in tensor networks. Tensor networks are one of the most efficient classical frameworks to simulate quantum mechanical systems \cite{biamonte2017tensornetworksnutshell}. This approach has been recently included in Qiskit functions \cite{Qiskit_TEM}.

\begin{figure}[t]
    \centering
    \resizebox{\columnwidth}{!}{%
    \input{circuits/tem_circuit}
    }
    \caption{Quantum-classical circuit implements an example of a tensor network error mitigation algorithm. The operations $U_i$ are gates and $\mathcal{N}_i$ are quantum channels that model the noise.}
    \label{fig:tem_circuit}
\end{figure}
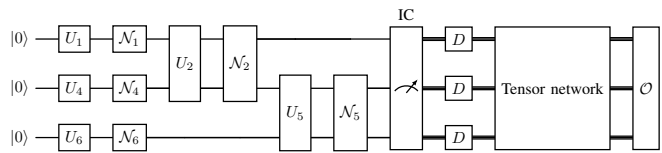

Tensor network error mitigation consists of three major components. Initially, the goal is to execute a quantum circuit on noisy hardware and estimate the expectation value of an observable as described in \autoref{eq:expectation_value}. Instead of estimating the expectation value from the noisy quantum computer, the tensor network error mitigation algorithm inserts a noise-mitigating layer between the hardware circuit and the observable. This layer comprises an informationally complete POVM measurement followed by a tensor network that applies operations that aim to cancel error. The algorithm prepares an IC-POVM measurement based on the measurement operators in \autoref{eq:common_single_qubit_povms} and reconstructs the state using the dual operators in \autoref{eq:dual_operator} and formulation in \autoref{eq:IC_POVM_reconstructed_state}. After rebuilding the noisy state with an IC-POVM, the state works as an input for a tensor network, which applies the quantum channels that invert the effects of noise on the hardware. The algorithm is summarized in \autoref{fig:tem_circuit}.

%% file: tables/wire_cutting_example.tex
\begin{table}[t]
\centering
\begin{tabular}{c|c|c|c}
\textbf{$i$} & \textbf{$P_i$} & \textbf{$\rho_i$} & \textbf{$c_i$} \\
\hline
1 & $I$ & $|0\rangle \langle 0|$ & $+1/2$ \\
2 & $I$ & $|1\rangle \langle 1|$ & $+1/2$ \\
3 & $X$ & $|+\rangle \langle +|$ & $+1/2$ \\
4 & $X$ & $|-\rangle \langle -|$ & $-1/2$ \\
5 & $Y$ & $|+i\rangle \langle +i|$ & $+1/2$ \\
6 & $Y$ & $|-i\rangle \langle -i|$ & $-1/2$ \\
7 & $Z$ & $|0\rangle \langle 0|$ & $+1/2$ \\
8 & $Z$ & $|1\rangle \langle 1|$ & $-1/2$ \\
\end{tabular}
\vspace{1em}
\caption{Measurements $P_i$, associated states $\rho_i$, and coefficients $c_i$.}
\label{table:wire_cutting}
\end{table}

%% file: circuits/wire_cutting_example.tex
\begin{quantikz}
   \qw \gategroup[3,steps=5,style={rounded corners,fill=yellow!30,opacity=0.5,inner ysep=1pt}, background]{} & \gate[2]{} & \qw & \qw & \qw \\
   \qw & \qw & \push{\textbf{CUT}} & \gate[2]{}  & \qw \\
   \qw & \qw &  \qw & \qw & \qw
\end{quantikz} $\Longrightarrow$
\begin{quantikz}
   \qw \gategroup[2,steps=6,style={rounded corners,fill=blue!30,opacity=0.5,inner ysep=1pt}, background]{} & \gate[2]{} & \qw & \qw & \qw & \qw \\
    & \qw  & \qw & \gate{P_i} & \wireoverride{n} \\ 
   \qw \gategroup[2,steps=6,style={rounded corners,fill=red!30,opacity=0.5,inner ysep=1pt}, background]{} & \wireoverride{n}  & \wireoverride{n}  & \gate{|\rho_i \rangle} \wireoverride{n} & \gate[2]{} & \qw \\
    & \qw & \qw & \qw & \qw & \qw
\end{quantikz}

%% file: circuits/wire_cutting_identity.tex
\begin{quantikz}
\lstick{$\rho$} & \gate{I} & \rstick{$\rho$} \qw
\end{quantikz}
\quad = \quad \sum_{i=1}^{8} c_i
\begin{quantikz}
\lstick{$\rho$} & \gate{P_i} & \wireoverride{n} & \lstick{\ket{\rho_i}} \wireoverride{n} & \rstick{$\rho$}
\end{quantikz}

%% file: circuits/gate_cutting_example.tex
\begin{quantikz}
 \qw \gategroup[2,steps=6,style={rounded corners,fill=blue!30,opacity=0.5,inner ysep=1pt}, background]{} & \gate[2]{} & \gate[1]{} & \gate[3, style={fill=green!30}]{} & \qw & \qw \\
 \qw & \qw & \gate[3, style={fill=green!30}]{} & \qw & \qw & \qw \\
 \qw \gategroup[2,steps=6,style={rounded corners,fill=red!30,opacity=0.5,inner ysep=1pt}, background]{} & \qw & \qw & \qw & \gate[2]{} & \qw \\
 \qw & \qw & \qw & \gate[1]{} & \qw & \qw
\end{quantikz} $\Longrightarrow$
\begin{quantikz}
\qw \gategroup[2,steps=6,style={rounded corners,fill=blue!30,opacity=0.5,inner ysep=1pt}, background]{} & \gate[2]{} & \gate[1]{} & \gate[2, style={fill=green!30}]{} & \qw & \qw \\
 \qw & \qw & \gate[1, style={fill=green!30}]{} & \qw & \qw & \qw \\
 \qw \gategroup[2,steps=6,style={rounded corners,fill=red!30,opacity=0.5,inner ysep=1pt}, background]{} & \qw & \gate[2, style={fill=green!30}]{} & \gate[1, style={fill=green!30}]{} \qw & \gate[2]{} & \qw \\
 \qw & \qw & \qw & \gate[1]{} & \qw & \qw
\end{quantikz}

%% file: tables/gate_cutting_example.tex
\begin{table}[t]
\centering
\begin{tabular}{c|c|c}
\textbf{$i$} & \textbf{$a_i$} & \textbf{$\mathcal{F}_i$} \\
\hline
1 & $\cos^2(\theta/2)$ & $I \otimes I$ \\
2 & $\sin^2(\theta/2)$ & $Z \otimes Z$ \\
3 & $-\sin^2(\theta/2)$ & $P_z \otimes S$ \\
4 & $\sin^2(\theta/2)$ & $P_z \otimes S^{\dagger}$ \\
5 & $-\sin^2(\theta/2)$ & $S \otimes P_z$ \\
6 & $\sin^2(\theta/2)$ & $S^{\dagger} \otimes P_z$ \\
\end{tabular}
\vspace{1em}
\caption{Coefficients $a_i$ and corresponding channels $\mathcal{F}_i$.}
\label{table:gate_cutting_example}
\end{table}

%% file: circuits/tem_circuit.tex
\begin{quantikz}
\lstick{$\ket{0}$} & \gate{U_1} & \gate{\mathcal{N}_1} & \gate[2]{U_2} & \gate[2]{\mathcal{N}_2} & \qw & \qw & \meter[3]{\text{IC}} & \gate{D} \cw & \gate[3]{\text{Tensor network}} \cw & \gate[3]{\mathcal{O}} \cw \\
\lstick{$\ket{0}$} & \gate{U_4} & \gate{\mathcal{N}_4} & \qw & \qw & \gate[2]{U_5} & \gate[2]{\mathcal{N}_5} & \qw & \gate{D} \cw & \cw & \cw \\
\lstick{$\ket{0}$} & \gate{U_6} & \gate{\mathcal{N}_6} & \qw & \qw & \qw & \qw & \qw & \gate{D} \cw & \cw & \cw \\
\end{quantikz}

%% file: sections/conclusion.tex
\section{Conclusion}

This article reviewed the measurement operations in quantum computing from an algorithmic perspective. First, we presented the key definitions in quantum measurement theory, which contained projective measurements and POVM measurements. We divided the measurement applications into three categories: measurements in static circuits, measurements in dynamic circuits, and measurements for various quantum computational challenges. We discussed the role of measurements in variational and fundamental algorithms. We showed how measurements are essential in quantum state preparation and quantum teleportation. Finally, we reviewed measurement schemes that are used in circuit cutting and error mitigation. 

Measuring creates challenges: it introduces statistical and hardware-related errors in the algorithms. On the other hand, a proper set of measurements can reveal much more about the state than the standard measurement on a computational basis. Mid-circuit measurements can be used to reduce circuit depth and implement novel distributed quantum computing algorithms. Moreover, measurements will be key components in fault-tolerant quantum computing. We believe that novel measurement schemes will play increasingly crucial roles in future quantum algorithms.

Ultimately, measurement is not just a passive readout method but an active component shaping the computational capabilities of quantum devices. Understanding its full potential will be essential in realizing error-prone and practically useful quantum computing.

\section*{Acknowledgment}

The authors would like to thank Abigail Parker for valuable comments and suggestions that helped improve the quality of this paper.